\documentclass{aastex631}

\usepackage{bm}
\usepackage{xcolor}
\usepackage{multibib}
\newcites{main}{References}
\newcites{appendix}{References (Appendix)}
\usepackage{subfigure}
\usepackage{soul}
\usepackage{tcolorbox}
\usepackage{amsmath}
\usepackage{verbatim}
\newcommand{\beq}{\begin{equation}}
\newcommand{\eeq}{\end{equation}}
\newcommand{\bseq}{\begin{subequations}}
\newcommand{\eseq}{\end{subequations}}
\newcommand{\bary}{\begin{eqnarray}}
\newcommand{\eary}{\end{eqnarray}}
\newcommand{\bwt}{\begin{widetext}}
\newcommand{\ewt}{\end{widetext}}

\begin{document}
\title{
A photohadronic interpretation of the H.E.S.S. afterglow observations from GRB 221009A 
}
\author[0000-0003-0038-5548]{Sarira Sahu}
\email{sarira@nucleares.unam.mx}
\affiliation{Instituto de Ciencias Nucleares, Universidad Nacional Aut\'onoma de M\'exico,\\ Circuito Exterior                  S/N, C.U., A. Postal 70-543, CDMX 04510, México.}

\author[0009-0009-4471-7242]{B. Medina-Carrillo}
\email{benjamin.medina@cinvestav.mx}
\affiliation{Departamento de Física Aplicada, Centro de Investigación y de Estudios Avanzados del IPN,\\ Unidad Mérida. A. Postal 73, Cordemex, Mérida, Yucatán 97310, México.}

\author[0009-0001-2172-4556]{D. I. Páez-Sánchez}
\email{paez@ciencias.unam.mx}
\affiliation{Facultad de Ciencias, Universidad Nacional Aut\'onoma de M\'exico, \\ 
Circuito Exterior, C.U., A. Postal 70-543, 04510 CDMX, México.}

\author[0000-0002-0291-2412]{G. Sánchez-Colón}
\email{gabriel.sanchez@cinvestav.mx}
\affiliation{Departamento de Física Aplicada, Centro de Investigación y de Estudios Avanzados del IPN,\\ Unidad Mérida. A. Postal 73, Cordemex, Mérida, Yucatán 97310, México.
}

\author[0000-0003-1967-7217]{Subhash Rajpoot}
\email{Subhash.Rajpoot@csulb.edu}
\affiliation{Department of Physics and Astronomy, California State University,\\
1250 Bellflower Boulevard, Long Beach, CA 90840, USA.
}

\begin{abstract}
The High Energy Stereoscopic System (H.E.S.S.) started observing the extremely powerful long-duration gamma-ray burst, GRB 221009A, after 53 hours of the triggering event. The H.E.S.S. collaboration carried out observations on the 11, 12, and 17, of October 2022 under poor atmospheric conditions, without detecting significant very high-energy photons from the source and computed the upper limits of the fluxes for the different nights. We study these flux upper limits by using the photohadronic model and show that the interaction of high-energy protons with the synchrotron seed photons in the forward shock region of the GRB jet exhibits behavior compatible with the upper limits computed by the H.E.S.S. collaboration. 

\end{abstract}


\section{Introduction}

The long-duration gamma-ray burst, GRB 221009A, was first observed by the Gamma-ray Burst Monitor~\citep{Meegan_2009} onboard the Fermi Gamma-ray Space Telescope on October 9, 2022~\citep{2022GCN.32642....1L, 2022GCN.32636....1V}. The prompt emission was also observed by a fleet of space observatories~\citep{2022GCN.32660....1G, 2022GCN.32805....1K, 2022GCN.32663....1L, 2022GCN.32751....1L, 2022GCN.32650....1U, 2022GCN.32661....1X, 2022GCN.32668....1F, 2022GCN.32632....1D, 2022GCN.32688....1K, 2022GCN.32657....1P, 2022GCN.32746....1M, 2023arXiv230301203A,Ripa:2023ssj, Williams:2023sfk}. This is also the first James Webb Space Telescope (JWST) observation of a GRB afterglow which suggests that the extreme properties of GRB 22109A are not directly related to its galactic environment~\citep{Levan:2023doz}. The redshift of the GRB is measured to be $z\simeq 0.151$,  corresponding to a luminosity distance of $\sim 750$ Mpc~\citep{2022GCN.32648....1D, 2022GCN.32686....1C}. This is the brightest GRB ever detected with the highest isotropic equivalent total energy. For this reason it is refereed to as the brightest of all time (BOAT) event~\citep{ Burns:2023oxn}. Observation of such a titanic GRB at a closer distance is a very rare event~\citep{Williams:2023sfk, 2023arXiv230207891M}. Estimates show that such rare events occur, on average, once every 10,000 years~\citep{Burns:2023oxn}. 

The  Large High Altitude Air Shower Observatory (LHAASO) with its two detectors, water Cherenkov detector array (WCDA) and larger air shower kilometer square area (KM2A) detector~\citep{LHAASO:2019qtb}, observed several thousands very high-energy (VHE, $>100$ GeV) photons within $T_0+2000$ s in the 500 GeV to 18 TeV energy range, where $T_0$ is the trigger time of GRB 221009A. Again, these are the most energetic photons ever observed from a GRB~\citep{2022GCN.32677....1H}. The propagating VHE photons interact with the extragalactic background light (EBL) to produce $e^+e^-$ pairs ($\gamma\gamma\rightarrow e^+e^-$)~\citep{1992ApJ...390L..49S,Hauser:2001xs}, thus attenuating the energy-dependent VHE photon flux reaching the Earth. Several well known EBL models have been developed to study the attenuation at different redshifts and photon energies~\citep{Franceschini:2008tp,2010ApJ...712..238F, 2011MNRAS.410.2556D, 2012MNRAS.422.3189G, 2021MNRAS.507.5144S}. Particularly, the $\sim 18$ TeV photons from the GRB 221009A should be attenuated by a factor of $\sim 10^{-8}$,  making them almost impossible to observe on Earth. In spite of this, observations of such VHE photons from the GRB 221009A by LHAASO question the reliability of the EBL models used in the very high-energy regime. To overcome this problem, nonstandard physics explanations are proposed to explain the observation of $\sim 18$ TeV photons~\citep{2022arXiv221007172B,Zhu:2022usw,2022arXiv221005659G,Troitsky:2022xso,Brdar:2022rhc,Nakagawa:2022wwm,2023ApJ...942L..21F, Li:2022wxc}. Recent publication from the LHAASO collaboration has shown that GRB 221009A was observed during the prompt and the early afterglow emission phases in the VHE band by the WCDA. Photons of energy $> 0.2$ TeV were observed a few minutes after the triggering event. One way to understand the rise and fall of the VHE spectra at different epochs is entertained by using the leptonic scenario~\citep{2023Sci...380.1390L}.

The H.E.S.S. collaboration started observing the GRB 221009A with its 12 m telescopes after 53 hours of the triggering event when the observing condition were favourable~\citep{Aharonian_2023}. On October 11, 2022, the H.E.S.S. observation was for 32 minutes, followed by a second run of 32 minutes and continued observing the source in the subsequent nights. Due to poor atmospheric conditions and high aerosol content in the atmosphere~\citep{10.2307/27026830}, the only nights it could observe the source were on October 11, 12, and 17, of 2022 (Nights 3, 4, and 9, after the triggering event, respectively). Additional observations undertaken on other nights were excluded from the analysis due to the worsening of the atmospheric conditions~\citep{Aharonian_2023}. Unfortunately, H.E.S.S. did not detect significant emission of VHE gamma-rays from the source neither on the individual nights nor in the combined data. Instead, the collaboration computed the upper limits to constrain the VHE emission from GRB 221009A. In the energy range $0.65\, \mathrm{TeV} \le E_{\gamma} \le 10\, \mathrm{TeV}$, the computed  integrated flux upper limit  at 95\% confidence level (CL) for  Night 3 is $ 4.06\times 10^{-11 }\, \mathrm{erg \, cm^{-2 }\, s^{-1 }}$ and for all nights combined is $9.7\times 10^{-12 }\, \mathrm{erg \, cm^{-2 }\, s^{-1 }} $~\citep{Aharonian_2023}.

The relativistic jets of blazars and GRBs are pointing towards us and the emission mechanisms in these sources show many similarities, despite large differences in their masses, bulk Lorentz factors, and magnetic fields~\citep{Wang:2010nr,Nemmen:2012rd,Wu:2015opa}.The relativistic jets in AGNs and GRBs have many characteristics in common and have a similar energy dissipation efficiency~\citep{Nemmen:2012rd}. So, the particle acceleration mechanisms and the VHE emission mechanisms used in the context of blazars are also used to study the VHE emissions in the afterglow phases of GRBs. We have studied the multi-TeV flaring events from several blazars using the photohadronic model. Recently, we have applied the photohadronic model successfully to explain the VHE emission from the GRB 190114C, GRB 180720B and 190829A~\citep{Sahu:2020dsg,2022ApJ...929...70S}. Also, the observation of $\sim 18$ TeV photons from GRB 221009A are studied using the photohadronic model~\citep{Sahu:2022gvx}.

Our aim here is to provide an alternative interpretation to the upper limits of the VHE spectrum as observed by H.E.S.S.  We do this in the context of the photohadronic model and also analyze whether the seed photons in the synchrotron regime or in the synchrotron self-Compton (SSC) regime can be responsible for the results of these VHE afterglow observations.

The plan of the paper is as follows. In Section~\ref{Section2}, we review the photohadronic model and its  application to explain the VHE flaring events from high-energy peaked BL Lac objects (HBLs). In Section~\ref{Section3}, the common features of blazars and GRBs are discussed. Finally, we discuss the results and conclusions in Section~\ref{Section4}.

\section{The Photohadronic Model}\label{Section2}

Blazars belong to a subclass of active galactic nuclei (AGN)~\citep{1995PASP..107..803U,Romero:2016hjn}. They have non-thermal spectra produced by the relativistic jet oriented very close to the observer's line of sight~\citep{1995PASP..107..803U, VERITAS:2010vjk}. Rapid variability is a common feature in their entire electromagnetic spectrum. The spectral energy distributions (SEDs) of these blazars have a double peak structure~\citep{Abdo_2010}. The low-energy peak is due to the synchrotron radiation of low-energy electrons in the jet's magnetic field. The second peak is from the Compton scattering off the high-energy electrons with the self-produced synchrotron photons (synchrotron self Compton, i.e., SSC)~\citep{1992ApJ...397L...5M,1993ApJ...416..458D, 1994ApJ...421..153S,Murase_2012,Gao:2012sq}. To explain the VHE events from an HBL, a double jet structure scenario is assumed~\citep{Sahu:2019lwj,Sahu_2019}. In this scenario, the inner jet is smaller with a comoving radius $R'_f$ (the notation $^{\prime}$ implies quantity in comoving frame) and is enclosed by a larger cone of size $R'_b$ ($R'_f < R'_b$). Both the jets are moving with a bulk Lorentz factor $\Gamma$ and a Doppler factor ${\cal  D}$~\citep{Ghisellini:1998it,Krawczynski:2003fq}. For HBLs, $\Gamma\simeq {\cal D}$. The photon density in the outer jet region is $n'_{\gamma}$, which is estimated from the observed flux. On the other hand, the photon density in the inner jet region $n'_{\gamma, f}$ is unknown. Adiabatic expansion of the inner jet to the outer jet decreases the photon density in the inner region. To connect the inner and the outer jet regions, we proposed a scaling behavior of the photon densities as~\citep{Sahu:2015tua}
\begin{linenomath*}
\beq
\frac{n'_{\gamma, f}(\epsilon_{\gamma_1})}
{n'_{\gamma, f}(\epsilon_{\gamma_2})}\simeq \frac{n'_\gamma(\epsilon_{\gamma_1})}
{n'_\gamma(\epsilon_{\gamma_2})},
\label{densityratio}
\eeq
\end{linenomath*}
where $\epsilon_{\gamma}$ is the seed photon energy. In this equation, the left hand side is unknown and the right hand side is known. We use the above relation to express the unknown photon density in the inner region in terms of the known photon density in the outer region. In the inner jet region, the Fermi accelerated protons have a power-law spectrum given by
\begin{linenomath*}
\beq
\frac{dN}{dE_p} \propto E^{-\alpha}_p,
\eeq
\end{linenomath*}
where $E_p$ is the proton energy and $\alpha\ge 2$ is the spectral index for proton~\citep{1993ApJ...416..458D}. The value of $\alpha$ is different for non-relativistic shocks, highly relativistic shocks and oblique relativistic shocks~\citep{2005PhRvL..94k1102K,2012ApJ...745...63S}. The interaction of high-energy protons in the inner jet region produces the $\Delta$-resonance ($p\gamma\rightarrow \Delta^+$), and subsequently, the decay of the $\Delta$-resonance to $\gamma$-rays and neutrinos takes place by the decay of the intermediate $\pi^0$ and $\pi^+$ respectively~\citep{Sahu:2012wv}. Here we are concerned with the observation of $\gamma$-rays from the $\pi^0$ decay. The kinematical conditions for the $\Delta^+$-resonance process give
\begin{linenomath*}
\beq
E_{\gamma} \epsilon_{\gamma} \simeq 0.032 \frac{{\cal D}^2}{(1+z)^2}\, \mathrm{GeV}^2\textcolor{red}{.}
\label{KinCon}
\eeq
\end{linenomath*}
In the above process, the VHE photon carries about 10\% of the proton energy ($E_{p} \simeq 10E_{\gamma}$).

The interaction of the VHE photons from extragalactic sources with the EBL produces $e^+e^-$ pairs \citep{1992ApJ...390L..49S,2012Sci...338.1190A,Padovani:2017zpf}. Thus, the VHE photon flux is attenuated by a factor $e^{-\tau_{\gamma\gamma}}$, where the optical depth $\tau_{\gamma\gamma}$ depends on $E_{\gamma}$ and $z$. In the photohadronic model, the observed VHE $\gamma$-ray flux is determined to be
\begin{linenomath*}
\beq
F_{\gamma}(E_{\gamma})=F_0 \left ( \frac{E_\gamma}{\rm TeV} \right )^{-\delta+3}\,e^{-\tau_{\gamma\gamma}(E_\gamma,z)}=F_{in}(E_{\gamma})\, e^{-\tau_{\gamma\gamma}(E_{\gamma},z)},
\label{FluxObs}
\eeq
\end{linenomath*}
where the normalization factor $F_0$ can be fixed from the observed spectrum. The photon spectral index, $\delta=\alpha+\beta$, is the free parameter in the model~\citep{Sahu:2019lwj, Sahu_2019}. $F_{in}(E_\gamma)$ is the intrinsic flux and $\beta$ is the spectral index of the background seed photons. It is further observed that for HBLs the seed photon flux is also a power-law $\Phi_{\gamma}\propto  \epsilon^{\beta}_{\gamma}\propto E^{-\beta}_{\gamma}$. The sign of $\beta$ decides whether the seed photon background is in the synchrotron regime ($\beta < 0$) or in the SSC regime ($\beta > 0$). By fitting the observed VHE spectrum, the value of $\delta$ is fixed. For HBLs, we have shown that the value of $\delta$ is always in the range $2.5\le \delta \le 3.0$~\citep{Sahu_2019}. This makes the $\beta$ value to be positive.

\section{Common features of Blazars and GRBs}\label{Section3}

The emission mechanisms in blazars and GRBs have many common features despite enormous differences in their masses, bulk Lorentz factors, duration of emission and other factors~\citep{1995PASP..107..803U, 2011ApJ...740L..21W, Wang:2010nr, Nemmen:2012rd,2013FrPhy...8..661G, Wu:2015opa,BZhang}. In the past, we have studied very successfully the VHE gamma-ray flaring mechanism in high-energy blazars (HBLs) using the photohadronic model~\citep{Sahu:2019lwj,Sahu_2019}. Also, recently, the same photohadronic model was employed successfully to explain the VHE gamma-ray spectra of the GRB 180720B, GRB 190114C and GRB 190829A~\citep{Sahu:2020dsg,2022ApJ...929...70S}. Furthermore, the same model was employed to explain the observations of $E_{\gamma} \gtrsim 10$ TeV photons from the GRB 221009A~\citep{Sahu:2022gvx}. Due to the extreme nature of the GRB 22109A, it is assumed that during the afterglow epoch, the intrinsic photon flux from the source might have increased by several orders of magnitude\textcolor{red}{,} which might help to compensate the depletion from the $e^+e^-$ pair production of the VHE photons with the EBL. 
Previously, it is shown that for GRBs, the $\beta$ value can be either positive or negative~\citep{Sahu:2020dsg}. The seed photons in the synchrotron regime corresponds to $\beta < 0$~\citep{Sahu:2020dsg} and $\beta > 0$ corresponds to seed photons in the SSC regime~\citep{Sahu:2020dsg, 2022ApJ...929...70S}. However, for HBLs  $\beta > 0$ always and $2.5\le \delta \le 3.0$~\citep{Sahu:2019lwj}. In~\cite{Sahu:2022gvx}, it is demonstrated that the $E_{\gamma} \gtrsim 10$ TeV photons from the GRB 221009A can be explained in the context of the photohadronic process and $1.2 \lesssim \delta \lesssim 1.7$ is preferable. This range of $\delta$ implies that the spectral index of the background seed photon flux is negative and the seed photons are in the synchrotron regime in the forward shock region of the GRB jet.

\begin{figure}
\centering
\includegraphics[width=0.8\linewidth]{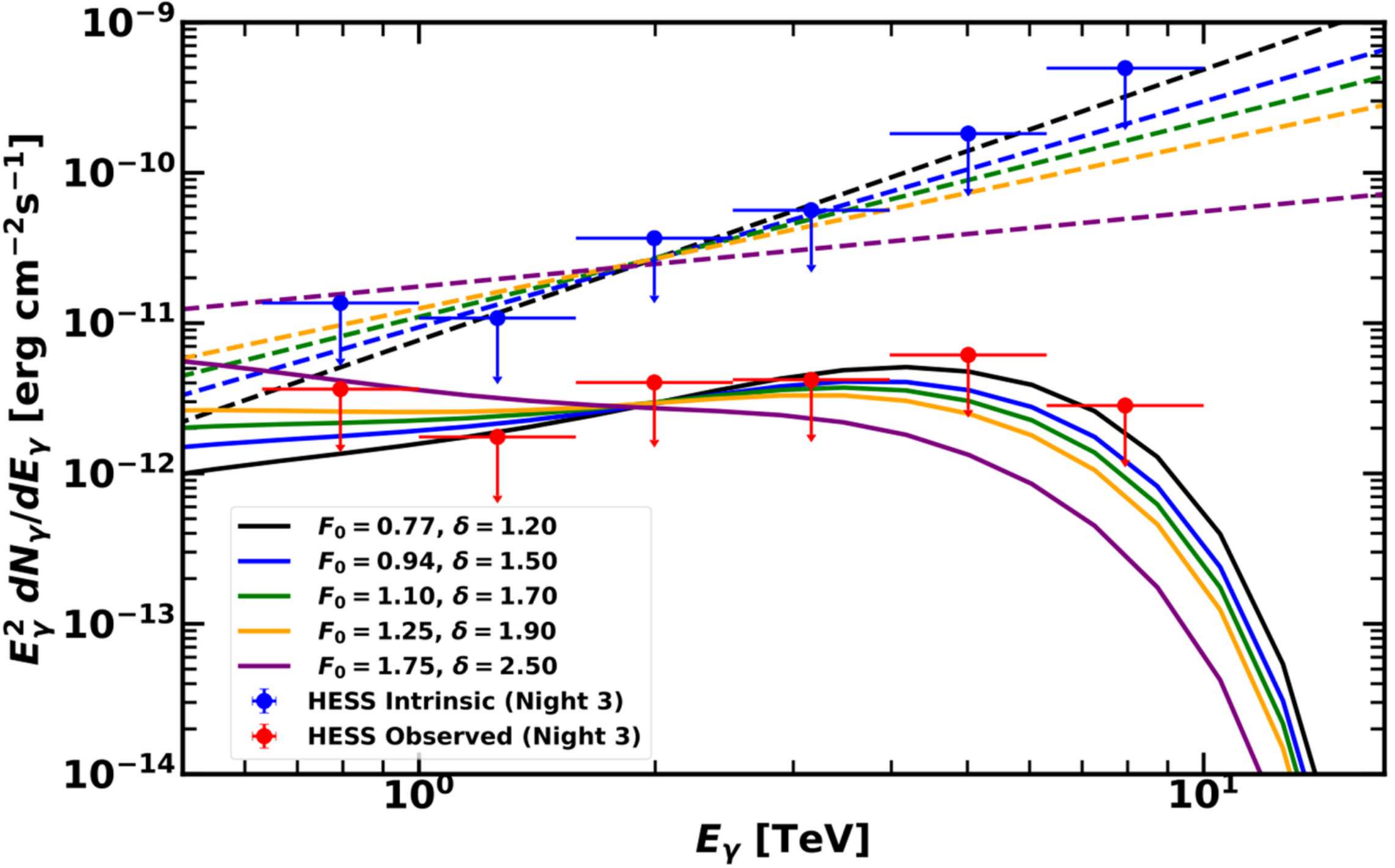}
\caption{ In the context of the photohadronic model, using different values of the spectral index $\delta$ and the normalization factor $F_0$ (in units of $10^{-11} \mathrm{erg\, cm^{-2}\, s^{-1}}$), we have plotted the calculated flux together with the H.E.S.S. upper limits on Night 3 (October 11, 2022). For every observed spectrum $F_\gamma(E_\gamma)$ (continuous curve), we have shown the corresponding intrinsic spectrum $F_{in}(E_\gamma)$ (dashed curve of the same color).}
\label{fig:figure1}
\end{figure}

\begin{figure}
\centering
\includegraphics[width=0.8\linewidth]{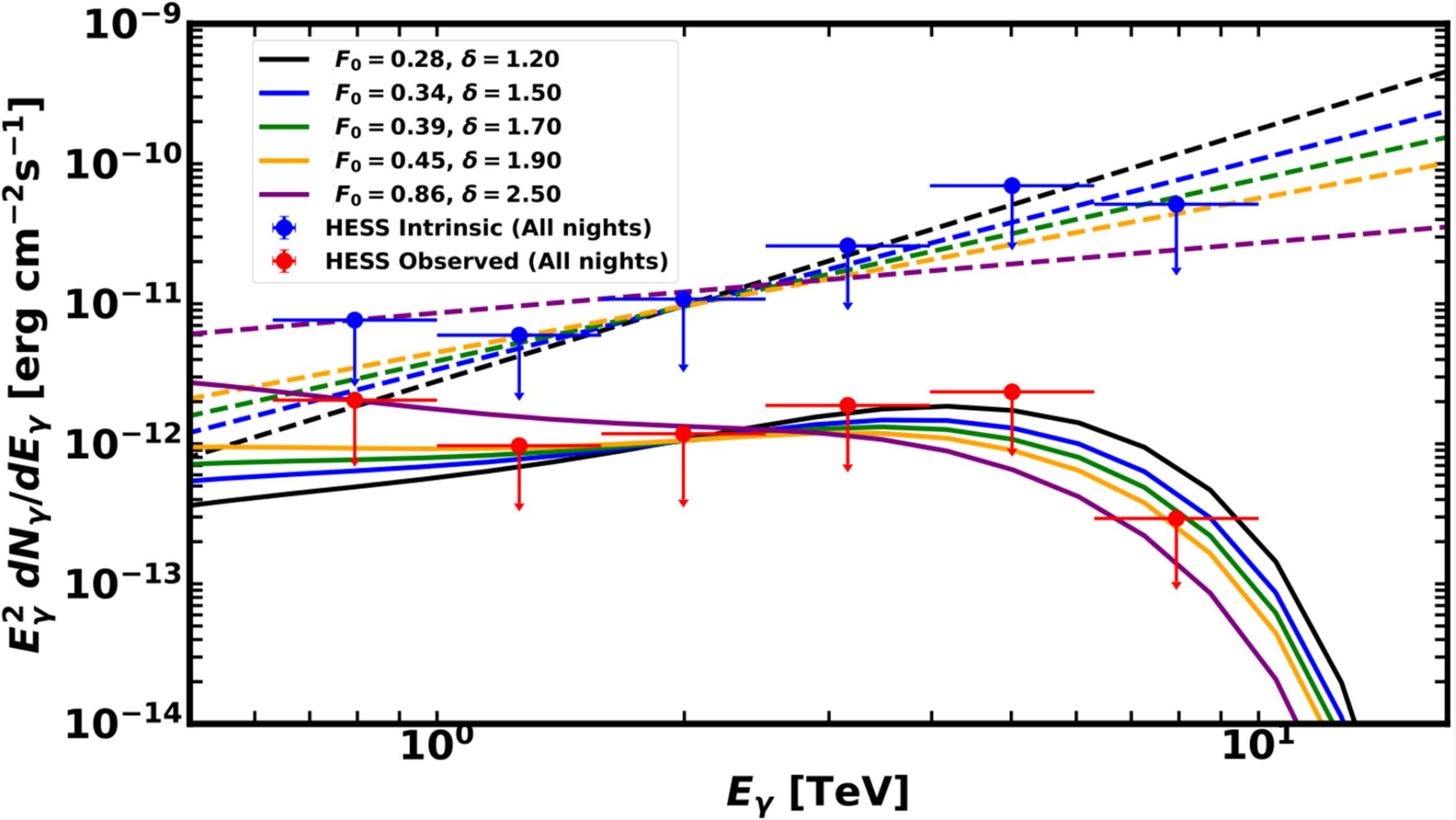}
\caption{In the context of the photohadronic model, using different values of the spectral index $\delta$ and the normalization factor $F_0$ (in units of $10^{-11} \mathrm{erg\, cm^{-2}\, s^{-1}}$), we have plotted the calculated flux together with the H.E.S.S. upper limits on All nights combined (Nights 3, 4 and 9). For every observed spectrum $F_\gamma(E_\gamma)$ (continuous curve), we have shown the corresponding intrinsic spectrum $F_{in}(E_\gamma)$  (dashed curve of the same color).
}
\label{fig:figure2}
\end{figure}

\begin{table}
\centering
\caption{For different values of $\delta$ we have calculated the normalization factor $F_0$ and the integrated flux $F^{int}_{\gamma}$ (in units of $10^{-11} \mathrm{erg\, cm^{-2}\, s^{-1}}$) in the energy range $0.65\, \mathrm{TeV} \le E_{\gamma}\le 10\, \mathrm{TeV}$ for Night 3 and for All nights combined. Corresponding to the integrated flux we have also estimated the luminosity $L_{\gamma}$ (in units of $10^{44}\, \mathrm{erg\, s^{-1}}$).}
\begin{tabular*}{\columnwidth}{@{\extracolsep{\fill}}lllll@{}}
\hline
$\delta$ &Interval& $F_0$ & $F^{int}_{\gamma}$ & $L_{\gamma}$ \\
\hline 
 1.2 &Night 3& 0.77 & 0.79 & 5.32 \\
 \, &All nights& 0.28& 0.29& 1.95\\
 \hline
 1.5 & Night 3&0.94 & 0.72& 4.84 \\
 \, & All nights&0.34& 0.26&1.75\\
 \hline
 1.7 & Night 3&1.10 & 0.69 & 4.64\\
 \,& All nights&0.39&0.25&1.68\\
 \hline
 1.9 &Night 3& 1.25  & 0.67& 4.51\\
 \, &All nights&0.45&0.24&1.61\\
 \hline
 2.5 &Night 3& 1.75& 0.64 & 4.31\\
 \,&All nights& 0.86& 0.32&2.15\\
\hline 
\end{tabular*}
\label{table1}
\end{table}

\section{Results and Conclusions}\label{Section4}

In spite of poor atmospheric conditions and high levels of moonlight, the H.E.S.S. collaboration estimated with 95\% CL, upper limits to the intrinsic fluxes on October 11 (Night 3) and the combined data set for the Nights of 3, 4, and 9. The striking feature of the estimated spectrum is that it increases slowly, peaks around 5 TeV and then decreases exponentially. By incorporating the EBL model of~\cite{Franceschini:2008tp} in the photohadronic model, we plotted the upper limits for Night 3 and for the combined fluxes of all the nights by using different values of $\delta$ and $F_0$. The results are shown in Figure~\ref{fig:figure1} and Figure \ref{fig:figure2}, respectively. Here, our analysis is based on the EBL model of~\cite{Franceschini:2008tp}. It is observed that the EBL model of~\cite{2011MNRAS.410.2556D} has similar results. In Figure~\ref{fig:figure1}, it is observed that for $1.2\lesssim \delta \lesssim 1.9$ the predicted spectrum is compatible with the H.E.S.S. upper limits. The intrinsic flux in this model is given by $F_{in}\propto E^{1.8-1.1}_{\gamma}$. Also, the estimated photohadronic model curves have peaks around 5 TeV, consistent with the H.E.S.S. estimate. However, for $\delta > 1.9$ (e.g. $\delta=2.5$), the flux has a large value in the low energy regime and decreases as the energy increases. Irrespective of the upper limits, this trend can only be explained by taking $1.2 \lesssim \delta\lesssim 1.9$ and by adjusting the normalization factor $F_0$. In this case, the flux slowly increases to a large value for $E_{\gamma} \sim 5$ TeV and then decreases. However, for $\delta \gtrsim 2.0$ the flux slowly decreases from a large value and is thus unable to explain the trend. From the above analysis, we conclude that large values of $\delta$ ($\delta\gtrsim 2.0$) are not compatible with the H.E.S.S. results. We have also shown the upper limits to the intrinsic fluxes corresponding to each value of $\delta$ in the same figure. The same behavior is also observed for the flux from the combined data set, as shown in Figure~\ref{fig:figure2}. In the context of the photohadronic model, the VHE gamma-ray photons observed during the afterglow (within $T_0$+2000 s) by LHAASO and the upper limits reported from the 3rd to the 9th nights by H.E.S.S. are produced from the interactions of the Fermi accelerated high-energy protons with the photons in the descending part of the synchrotron spectrum in the forward shock region of the GRB jet. It is similar to the VHE gamma-ray flux observed during the afterglow from GRB 180720B~\citep{Sahu:2020dsg}. The GRB 221009A could produce multi-TeV gamma-rays for more than a week after the prompt phase. This feature is unique to GRB 221009A. In Table~\ref{table1} we have shown the normalization constant $F_0$, the integrated flux $F_{\gamma}^{int}$, and the luminosity $L_{\gamma}$ in the energy range $0.65\, \mathrm{TeV} \leq E_{\gamma} \le 10\, \mathrm{TeV}$ for the observations on Night 3 and the combined nights for different values of the spectral index $\delta$. The estimated $F_{\gamma}^{int}$ for each $\delta$ is smaller than the upper limit of H.E.S.S.~\citep{Aharonian_2023}.

The VHE photons in the energy range $0.65\, \mathrm{TeV} \le E_{\gamma} \le 10\, \mathrm{TeV}$ correspond to protons in the energy range $6.5\, \mathrm{TeV} \le E_p \le 100\, \mathrm{TeV}$. These protons are accompanied by the electrons in the same energy range. Both the electrons and the protons radiate synchrotron photons in the GRB jet ~\citep{1991A&A...251..723M,2001APh....15..121M,2003APh....18..593M,2008PhRvD..78c4013K}. The synchrotron photons radiated by the protons will be suppressed by a factor of $m^{-4}_p$, where $m_p$ is the proton mass, relative to the synchrotron photons from the electrons. If the jet has a magnetic field of $\sim 1$ G, then the synchrotron photons from the electrons will be radiated in the energy range $2.2\, \mathrm{MeV} \le \epsilon_{\gamma} \le 522\, \mathrm{MeV}$ which falls in the lower part of the SSC spectrum. However, for smaller magnetic fields, this range of $\epsilon_{\gamma}$ will be reduced. 

If VHE photons were to interact with the seed photons in the GRB jet to produce $e^+e^-$ pairs~\citep{refId0}, then each lepton will have maximum energy of $E_{\gamma}/2$. Similarly, the $e^+$ produced from the $\pi^+$ decay will have energy in the range $0.33\, \mathrm{TeV} \le E_{e} \le 5\, \mathrm{TeV}$. Thus, in the jet magnetic field, the $e^+$ and $e^-$ will emit synchrotron photons in the energy range $6\, \mathrm{keV} \le \epsilon_{\gamma} \le 1.3\, \mathrm{MeV}$.

The GeV-TeV photons produced from the $\pi^0$ decay will mostly encounter the seed photons with energy $\epsilon_{\gamma} \geq 6$ keV as discussed above. The $e^+e^-$ pair production cross section for $\epsilon_{\gamma}\geq6$ keV is $\sigma_{\gamma\gamma}\lesssim  10^{-29}\, \mathrm{cm^{2}}$. After 53 hours of the triggering event of the GRB 221009A, the jet has a radius $R'_b\sim (10^{17} - 10^{18})$ cm. The mean free path for the $e^+e^-$ process is $\lambda_{\gamma\gamma}=(n'_{\gamma}\sigma_{\gamma\gamma})^{-1}$. For $\lambda_{\gamma\gamma} \le R'_b$ we need the photon density in the jet to satisfy $n'_{\gamma} \ge 10^{11}\, \mathrm{cm^{-3}}$ and such high photon density is difficult to achieve far away from the central engine. Thus, the VHE photons will not be attenuated in the jet environment. 

Another process, the Bethe–Heitler (BH) pair production process $p\,\gamma\rightarrow$ $p\,e^{+}e^{-}$, can also compete with the photohadronic process. The produced $e^+e^-$ pair can also emit synchrotron photons~\citep{Petropoulou:2014rla}. The BH process may be important for very high-energy protons and electrons, but here their contribution is negligible in the present context~\citep{Petropoulou:2015upa}. Thus, we conclude that the proton synchrotron process and the synchrotron process from the electrons and positrons will not affect the propagation of the VHE photons in the GRB jet environment.

We conclude that, in the context of the photohadronic scenario, the H.E.S.S. upper limits on the VHE flux in the energy range $0.65\, \mathrm{TeV} \leq E_{\gamma} \le 10\, \mathrm{TeV}$ are compatible, provided the high-energy protons in the GRB jet interact with the descending part of the synchroton spectrum and not with the SSC photons. Future observations of more GRBs in VHE gamma-rays are required to establish the validity of our claim.

We thank the referee for corrections/suggestions and constructive remarks that have considerably improved the manuscript. The work of S.S. is partially supported by DGAPA-UNAM (México) Projects No. IN103522. B. M-C and G. S-C would like to thank CONACyT (México) for partial support. Partial support from CSU-Long Beach is gratefully acknowledged. S.S. is thankful to J. D. Mbarubucyeye of the H.E.S.S. collaboration for fruitful discussions.


\bibliography{grbref}{}
\bibliographystyle{aasjournal}

\end{document}